\documentclass[aps,prl,reprint,superscriptaddress]{revtex4-1}

\usepackage{graphicx}
\usepackage{amssymb}
\usepackage{amsmath}

\begin{document}


\title{Nanoscale microwave imaging with a single electron spin in diamond}





\author{Patrick Appel}
\affiliation{Department of Physics, University of Basel, Klingelbergstrasse 82, Basel CH-4056, Switzerland}
\affiliation{These authors contributed equally}
\author{Marc Ganzhorn}
\affiliation{Department of Physics, University of Basel, Klingelbergstrasse 82, Basel CH-4056, Switzerland}
\affiliation{These authors contributed equally}
\author{Elke Neu}
\affiliation{Department of Physics, University of Basel, Klingelbergstrasse 82, Basel CH-4056, Switzerland}
\affiliation{Universitaet des Saarlandes, Experimentalphysik, D-66123 Saarbrücken, Germany}
\author{Patrick Maletinsky}
\affiliation{Department of Physics, University of Basel, Klingelbergstrasse 82, Basel CH-4056, Switzerland}


\date{\today}

\begin{abstract}
We report on imaging of microwave (MW) magnetic fields using a magnetometer based on the electron spin of a nitrogen vacancy center in diamond. We quantitatively image the magnetic field generated by high frequency (GHz) MW current with nanoscale resolution using a scanning probe technique. Together with a shot noise limited MW magnetic field sensitivity of $680$ nT/$\sqrt{\text{Hz}}$, our room temperature experiments establish the nitrogen vacancy center as a versatile and high performance tool for MW imaging, which furthermore offers polarization selectivity and broadband capabilities. As a first application of this scanning MW detector, we image the MW stray field around a stripline structure and thereby locally determine the MW current density with a  MW current sensitivity of a few nA/$\sqrt{\text{Hz}}$. 

\end{abstract}



\maketitle



Imaging and detecting MW fields constitutes a highly relevant element for engineering of future MW devices as well as for applications in atomic and solid state physics. For instance cavity quantum electrodynamics experiments with atoms~\cite{Kaluzny1983,Haroche2001} and superconducting qubits~\cite{Walraff2004,Nori2011} or the coherent control of quantum magnets~\cite{Thiele2014} and quantum dots~\cite{Koppens2006} are based on manipulating quantum systems with MW electric or magnetic fields. Precise control and knowledge of the spatial distribution of the MW near field is thereby essential to achieve optimal device performance. Also, magnetic systems are known to exhibit a large variety of collective magnetic excitations, including spin waves~\cite{Sparks1964} or excitations in frustrated magnets~\cite{Bishop2004,Balents2010}. Imaging such magnetic excitations on the nanoscale is a crucial step towards their fundamental understanding and the development of new spintronics devices, such as magnonic waveguides~\cite{Garcia2015} or domain wall racetrack memories~\cite{Parkin2008}. As a consequence, various techniques have been designed to image MW electric and magnetic fields, including scanning near field microscopy~\cite{Agrawal1997,Lee2000,Rosner2002}, micro-Brillouin light scattering~\cite{Sebastian2015}, superconducting quantum interference devices~\cite{Black1995} or imaging with atomic vapor cells~\cite{Boehi2012,Horsley2013,Affolderbach2015} or ultracold atoms~\cite{Boehi2010}. With only a few exceptions~\cite{Sebastian2015}, most of these techniques however lack a nanoscale spatial resolution or are restricted to operation in cryogenic or vacuum environments.

Microwave magnetic field imaging using the electronic spin of a single nitrogen vacancy (NV) center in diamond offers a promising alternative. The NV center is an optically active lattice point defect in diamond with a $S=1$ ground state manifold. Its atomic size, exceptionally long coherence times, optical initialization and readout of the spin state make the NV center an ideal sensor for DC magnetic fields under ambient conditions~\cite{Rondin2014,Degen2008,Taylor2008,Maze2008,Bala2008}. Recently, magnetometry of MW magnetic fields has been demonstrated using a NV spin in bulk diamond~\cite{Wang2015}, with a resulting MW magnetic field sensitivity of one $\mu$T/$\sqrt{\text{Hz}}$. However, the bulk NV centers employed in~\cite{Wang2015} severely restricted spatial resolution in imaging, and in particular do not allow for nanoscale imaging of MW near fields, which remains an outstanding challenge for NV-based MW imaging. In this letter, we address this issue and demonstrate the first nanoscale MW imaging using a scanning NV magnetometer~\cite{Maletinsky2012}. Our proof-of-concept imaging experiments were performed on a prototypical MW circuit - a micron-scale MW stripline - and yield nanoscale resolution combined with shot noise limited MW magnetic field sensitivites in the range of a few  hundred nT/$\sqrt{\text{Hz}}$.



\begin{figure*}[hbt!]
\includegraphics[width = 15.5 cm]{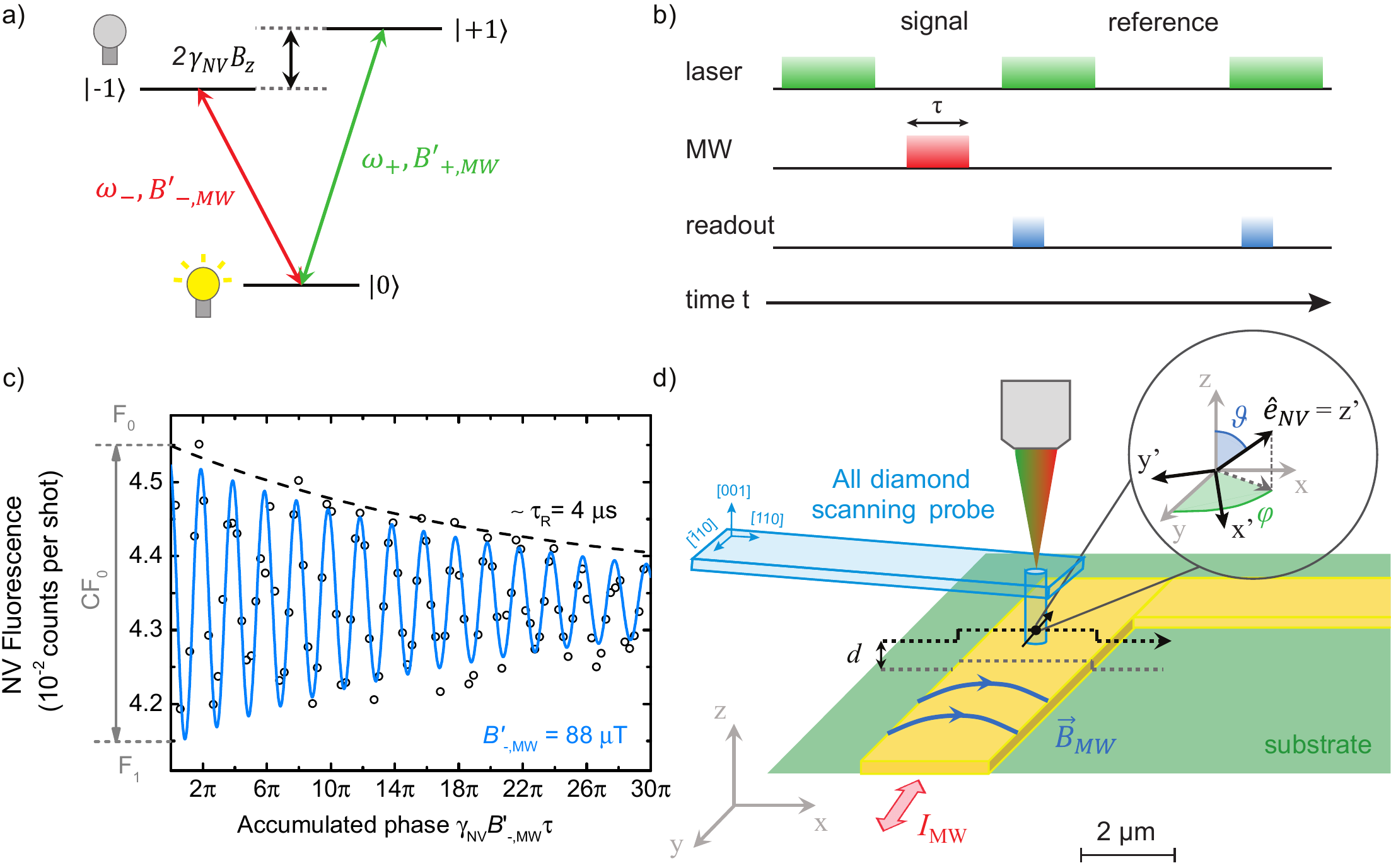}
\caption{
a) Energy levels of the NV spin in a static magnetic field $B_{\text{z}}$. The $|0\rangle$ state exhibits a higher fluorescence than the $|\pm 1\rangle$ states (bright and dark lightbulbs). 
b) Experimental pulse sequence employed for the measurement of NV Rabi oscillation. The first (second) readout pulse measures the fluorescence $F(B'_{-,\text{MW}})$ (bare fluorescence $F_0$) as described in the text. c) Optically detected Rabi oscillations of the NV spin, driven by a circular polarized MW magnetic field $B'_{-,\text{MW}}$ in a static magnetic field of $1.6$ mT ($\omega_{\text{MW}}/2\pi = 2.825$ GHz). The black dots are experimental data and the blue solid line corresponds to a fit with $F(\gamma_{\text{NV}} B'_{-,\text{MW}} \tau) = F_0(1-C/2) + C F_0/2 \cdot \cos\left({2 \pi \gamma_{\text{NV}} B'_{-,\text{MW}} \tau} \right) \cdot e^{-\tau / \tau_\text{R}}$, where $F_0$ is the fluorescence in the $|0\rangle$ state, $C$ is the amplitude and $\tau_{\rm R}$ the decay time of the Rabi oscillations. 
(d) Schematic representation of the combined confocal and atomic force microscope. The all-diamond scanning probe containing the NV spin is scanned at a height $d$ over the Pd stripline (yellow). The magnetic field generated by the current $I_{\text{MW}}$ which passes through the stripline is detected by the NV magnetometer. The NV spin is optically addressed using a homebuilt confocal microscope.}
\label{fig:fig1}
\end{figure*}

Our MW magnetic field detection is based on the ability of the MW field to drive coherent Rabi oscillations between the $|0\rangle$ and $|\pm 1 \rangle$ spin-states of the NV center (Fig. \ref{fig:fig1}a)-c)). Selection rules impose that within the rotating wave approximation (RWA), the transition $|0\rangle \rightarrow |\pm 1\rangle$ is only excited by a circularly polarized MW field $\sigma_{\pm}$. Due to the large NV spin splitting and the comparably weak microwave field amplitudes, the RWA holds to an extremely good extent in the experiments described here. An arbitrarily polarized MW field resonant with either the $|0\rangle \rightarrow |+1\rangle$ or $|0\rangle \rightarrow |-1\rangle$ transition therefore leads to an oscillation of the population between the two involved spin states, at a frequency $\Omega_{\pm}/2\pi=\gamma_{NV}B'_{\pm,MW}$, where $B'_{\pm,MW}$ is the (right-) left-handed circularly polarized component of the MW field in a plane perpendicular to the NV axis and $\gamma_{\text{NV}} = 28$ kHz/$\mu$T the NV gyromagnetic ratio. Measuring $\Omega_{\pm}$ by an appropriate experimental sequence (Fig. \ref{fig:fig1}b)) thus allows one to directly determine the amplitude of the driving MW magnetic field in a circularly polarized basis (Fig. \ref{fig:fig1}c)).

The NV spin we employ for MW imaging is located at the apex of an all diamond scanning probe (Fig.\ref{fig:fig1}d)), obtained in a series of fabrication steps, including low energy ion implantation, electron beam lithography and inductively coupled reactive ion etching~\cite{Maletinsky2012}. In order to perform MW magnetic field imaging, the diamond scanning probe is integrated in a homebuilt combined confocal (CFM) and atomic force microscope (AFM)~\cite{Maletinsky2012}. The AFM allows scanning of the NV spin in close proximity to a sample while the CFM is employed to optically read out the NV center spin state (Fig.\ref{fig:fig1}d)).  

We demonstrate the performance of our MW magnetic field imaging on the MW stripline structure illustrated in Fig. \ref{fig:fig1}d). The 2.5 $\mu$m wide MW stripline is patterned onto an undoped Si substrate covered with 300 nm of SiO$_2$ by electron beam lithography and evaporation of 60 nm of Pd. A MW source (Rhode\&Schwarz SMB 100A) is used to drive a MW current $I_{\text{MW}}$ with a frequency in the GHz range through the stripline. The right-angled stripline we employ thereby generates a highly inhomogeneous MW magnetic field with a nontrivial field distribution which is largely linearly polarized~\cite{SI}. This arrangement is therefore ideal to demonstrate spatial resolution and MW magnetic field sensitivity of our imager.

In the following, we demonstrate imaging of the $\sigma_-$-component of the MW magnetic field ($\omega_{\text{MW}}/2\pi = 2.825$ GHz) generated by the stripline. To that end, we tune the sensing frequency $\omega_{-}/2\pi = 2.87 $ GHz $ - \gamma_{\text{NV}} B_{\text{z}}$ of the $|0\rangle \rightarrow |- 1\rangle$ transition via a static magnetic field $B_z$ to the frequency $\omega_{\text{MW}}/2\pi$ of the MW field (see Fig. \ref{fig:fig1}a)). In analogy, the $\sigma_+$-component of the MW magnetic field can in principle be adressed by changing the external magnetic field $B_{\text{z}}$ such that the $|0\rangle \rightarrow |+ 1\rangle$ transition of the NV center is resonant with the frequency of the MW field to be imaged. However, we did not perform such an experiment, as for the device under test the MW magnetic field contains equal contributions of circular polarizations and would therefore not provide any polarization contrast in imaging~\cite{SI}. 


\begin{figure*}[hbt!]
\begin{center}
\includegraphics[width = 15.5 cm]{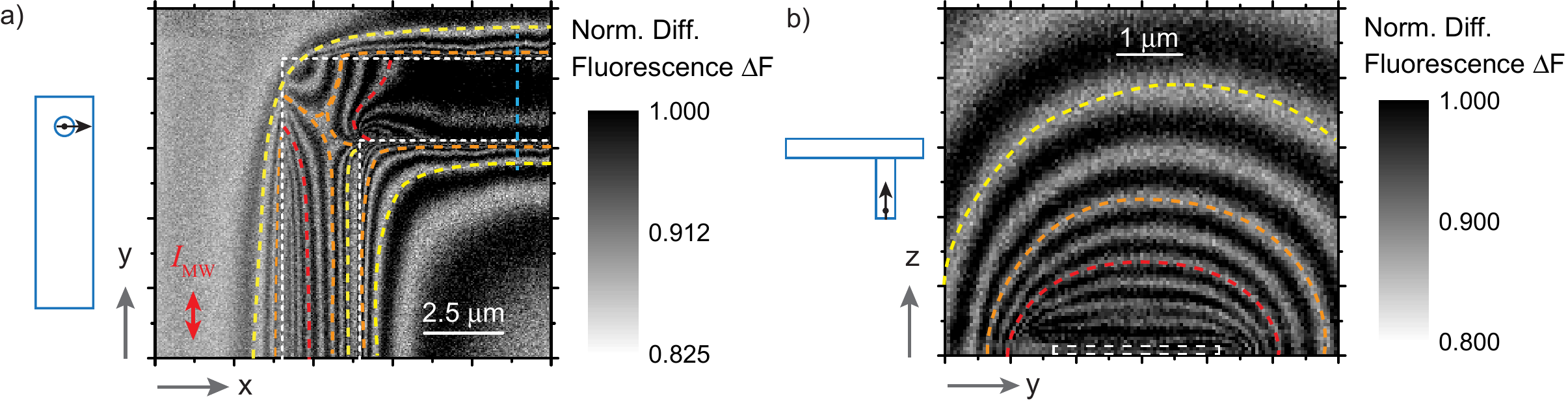}
\end{center}
\caption{Three dimensional isofield imaging of a MW magnetic field in 3D: 
Normalized differential fluorescence $\Delta F$ (see text) was acquired a) in the xy plane, at a distance $d$ above the sample and the stripline and b) in the zy plane which is indicated by the blue line in a.). The measurements were performed at a frequency $\omega_{\text{MW}}/2\pi = 2.825$ GHz with an input power at the stripline of $7.5$ dBm and $12.5$ dBm, respectively. The white dashed lines outline the stripline, whereas the dashed yellow, orange and red lines highlight reference isofield lines for $B'_{-,\text{MW}}= 360, 600$ and $840$ $\mu$T, respectively. The scanning probe is outlined in light blue (with the projection of the NV orientation into the respective plane depicted by the black arrow) and the MW current direction is labeled by a red arrow.}
\label{fig:fig2}
\end{figure*}

In the first imaging mode, we perform imaging of equi-magnetic field lines of constant amplitude $B'_{-,MW}$. For a fixed microwave pulse length $\tau_0$, the accumulated phase (pulse area) in the Rabi oscillation and thus the population difference between $|0\rangle$ and $|- 1\rangle$ depends on the local microwave magnetic field  $B'_{-,MW}$ (Fig. \ref{fig:fig1}b)-c)). While scanning the NV spin at a distance $d$ over the stripline, one can therefore monitor variations of the microwave magnetic field \textit{via} changes in the NV fluorescence $F$ (Fig. \ref{fig:fig2}).
In order to correct for fluorescence changes arising from potential near field effects~\cite{Tetienne2012,Novotny2006,Maletinsky2012} while scanning, we simultaneously record the bare fluorescence rate $F_0$ of the $|0\rangle$ state to yield the normalised differential fluorescence,  $\Delta F = \left[F\left(B'_{-,\text{MW}}\right)-F_0\right]/F_0$.

Figure \ref{fig:fig2}a) shows $\Delta F$ recorded in the xy plane in AFM contact (corresponding to a height $d$ of the NV spin above the stripline) with $\tau_{0} = 300$ ns. Each bright fringe corresponds to an integer multiple of $2\pi$ of accumulated phase of the NV Rabi oscillations. Consequently, the bright fringes represent isofield lines of $B'_{-,\text{MW}}$, which are spaced by $2\pi/\gamma_{\text{NV}}\tau_0=120~\mu$T. To avoid ambiguities in assigning the correct value of $B'_{-,\text{MW}}$ to each measured field line, we separately measured $B'_{-,\text{MW}}$ for several reference lines. The references for $B'_{-,\text{MW}}=360, 600$ and $840~\mu$T are highlighted in yellow, orange and red, respectively in Fig. \ref{fig:fig2}a).

The versatility and stability of our microscope allows us to further image MW magnetic fields in all three dimensions and in particular as a function of distance to the sample. To that end, we release AFM force feedback and record the MW magnetic field image by scanning the sample in a plane orthogonal to the MW current (Fig. \ref{fig:fig2}b)). In analogy to Fig. \ref{fig:fig2}a), we attribute a MW magnetic field amplitude to each isofield line as shown in Fig. \ref{fig:fig2}b).


\begin{figure*}[hbt!]
\includegraphics[width = 15.5 cm]{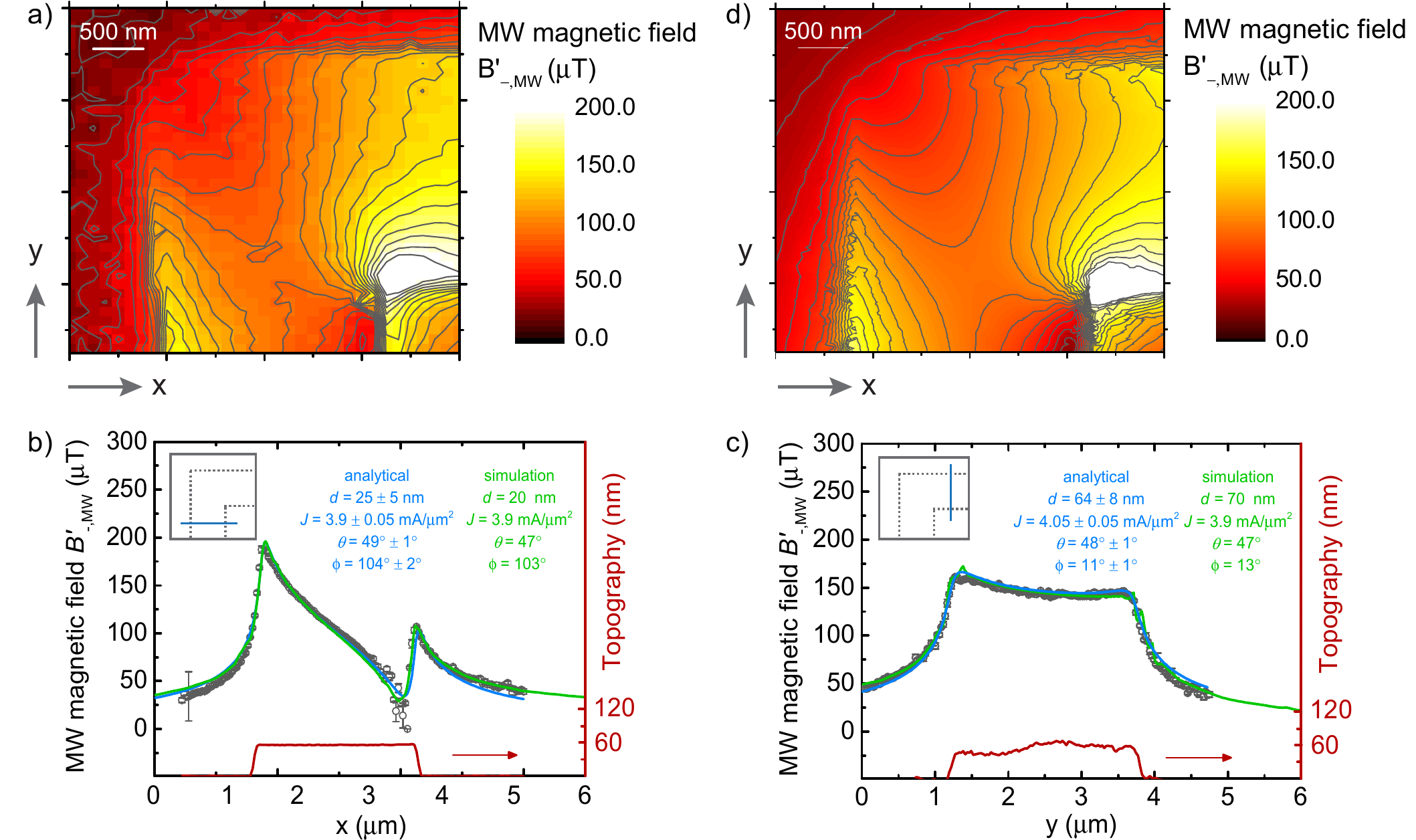}
\caption{Full, quantitative field mapping: a) Measured 2D spatial distribution of the MW magnetic field amplitude $B'_{-,\text{MW}}$. b),c) MW magnetic field $B'_{-,\text{MW}}$ (dark grey dots) and topography (red solid line) recorded during two different linescans. The inset shows the stripline outlined by the grey dashed line and the blue line depicts the direction of the respective linescan. Dark grey dots are the experimental data obtained from Rabi fits (see text) and the error bars correspond to the error of the Rabi fits. The blue lines correspond to the fit with the analytical function $B'_{-,\text{MW}}\left(d,J,\varphi,\theta\right)$ (see text), with the fit parameters and uncertainties~\cite{SI} being indicated in blue. The green lines are the fits with a finite element simulation (see text), with the fit parameters shown accordingly. The orientation of the scanning probe is the same as in Fig \ref{fig:fig2}b) and is identical for both linescans, resulting in a difference of $90^{\circ}$ for the azimuthal angle $\varphi$. The measurements have been performed at a frequency $\omega_{\text{MW}}/2\pi= 2.825$ GHz with a MW power of $7.5$ dBm at the stripline input. d) Simulated 2D spatial distribution of the MW magnetic field amplitude $B'_{-,\text{MW}}$ obtained by finite element computation of a 50 $\Omega$ stripline.}
\label{fig:fig3}
\end{figure*}

While providing a fast and straightforward method for nanoscale imaging of MW magnetic fields, our method for iso-field imaging suffers from limitations in regions of high magnetic field gradients. This is particularly appreciable near the edges of our stripline (Fig. \ref{fig:fig2}a)), where individual field lines are hard to distinguish and identification of the measured isofield lines becomes intractable. In order to overcome this limitation, we extended our imaging capabilities to directly determine $B'_{-,MW}$ at each point throughout the scan (Fig. \ref{fig:fig3}). For this, we measured NV Rabi oscillations at each pixel in the scan range and determined $B'_{-,\text{MW}}$ by a sinusodial fit to each of these traces. Fig. \ref{fig:fig3}a) depicts the resulting image of $B'_{-,MW}$ measured above the corner of the stripline imaged in Fig. \ref{fig:fig2}a). From this data, we also extract iso-field lines as highlighted by grey solid lines in Fig. \ref{fig:fig3}a). For the quantitative analysis of our results, which we provide below, we further used this imaging method to record linecuts of $B'_{-,\text{MW}}$ as depicted in Fig. \ref{fig:fig3}b) and c).

The measured distributions of $B'_{-,\text{MW}}$ depends on the orientation $\left(\varphi,\theta\right)$ and the position $\vec{r} = \left(x,y,z\right)$ of the NV spin with respect to the MW current~\cite{SI} (see also Fig. \ref{fig:fig1}). 
Assuming an infinitely thin stripline ($t \ll w$) in vacuum with a homogeneous MW current density $J$ oriented parallel to the stripline, the MW magnetic field profiles in Fig. \ref{fig:fig3}b) and c) can be described by an analytical function $B'_{-,\text{MW}}\left(d,J,\varphi,\theta\right)$~\cite{SI}, with $d,J,\varphi,\theta$ as free parameters. We note that our assumption of a homogenous current distribution is justified by the fact that the skin-depth of Pd at $2.825$ Ghz is larger than the stripline-width and is further corroborated by our numerical simulations~\cite{SI}. The resulting fits (blue lines in Fig. \ref{fig:fig3}b) and c)) are in excellent agreement with the experimental data (dark grey dots in Fig. \ref{fig:fig3}b) and c)). In addition, we have numerically computed in a finite element simulation the MW magnetic field amplitude, assuming a MW current in a stripline (width $w = 2.5$ $\mu$m, thickness $t = 60$ nm) on 300 nm of SiO$_2$~\cite{SI}. The best fit to the experimental data (green lines in Fig. \ref{fig:fig3}a) and b)) is achieved with parameters $d,J,\varphi,\theta$ (green insets) that are almost identical to the analytical fit parameters (blue insets). Finally, we also numerically determined the full two dimensional distribution of the MW magnetic field in a finite element simulation (Fig. \ref{fig:fig3}d)), using the distance and orientation of the NV spin determined in Fig. \ref{fig:fig3}b)~\cite{SI}. For most of the scanned area, the experimental data (Fig. \ref{fig:fig3}a)) are in excellent agreement with the simulation (Fig. \ref{fig:fig3}d)), which further establishes the reliability of our method.



The accurate determination of the NV-to-sample distance $d = 25 \pm 5$ nm that our method provides is relevant for various aspects of our work and NV-based sensing in general. First and foremost, $d$ determines the spatial resolution in imaging the sources of magnetic fields~\cite{Hingant2015,Maletinsky2012}, which we thus estimate to be $\sim25~$nm. 
Moreover, the distance links the MW current in the stripline to the MW magnetic field seen by the NV spin and therefore sets the sensitivity with which one can detect a MW current in the sample. With $d \sim 25$ nm and the magnetic field sensitivity determined below, we find a MW current sensitivity of our NV magnetometer of $\sim 300$ nA/$\sqrt{\text{Hz}}$ for an infinitely thin, current-carrying wire. Note that for the data set presented in Fig. \ref{fig:fig3}c), we find $d=64\pm 5~$nm, significantly larger than the value of $d=25\pm 5~$nm, which we determine for all the other data presented in this work. We attribute this discrepancy to contaminations on the diamond tip that has accumulated throughout the course of our experiments~\cite{SI} - removing these contaminants or working with a fresh tip should restore $d$ to it's original value. 

We now estimate the photon shot noise limited sensitivity of the NV spin determined by $\eta_{\text{photon}} = \sqrt{2e}/ \left(\pi \gamma_{\text{NV}} C \sqrt{F_0 \tau_{\text{R}}}\right)$, with $F_0$, $C$ and $\tau_R$ as defined earlier (see Fig. \ref{fig:fig1} and Ref.~\cite{SI}). For the NV spin used in our experiments, we find $\eta_{\text{photon}} = 680$ nT$/\sqrt{\text{Hz}}$ at 2.825 GHz~\cite{SI}. It should be noted that in general the decay time of the Rabi oscillation $\tau_R$ is itself a function of the Rabi frequency (and thus of $B'_{\text{MW}}$)~\cite{Slichter1996}. While a general expression for $\eta_{\text{photon}}$ is therefore difficult to obtain, it is instructive to consider the two limits of low and high Rabi frequencies with respect to $1/T_2^*$, where the decay time is given by $\tau_{\text{R}}=T_2^*$ and $T_1$, respectively. For typical values of NV centers in ultrapure diamond ($T_2^* \sim 1$ $\mu$s and $T_1 \sim 1$ ms) one then finds $\eta_{\text{photon}} \sim 1.4$ $\mu$T$/\sqrt{\text{Hz}}$ and $\eta_{\text{photon}} \sim 40$ nT$/\sqrt{\text{Hz}}$, respectively. Additionally, we note that while coherent detection of MW fields through Rabi oscillations is limited by $\tau_{\text{R}}$, incoherent detection of these fields is limited by $T_1$ only. Performing such incoherent MW imaging (also referred to as relaxation-imaging~\cite{Jayich2014,Tetienne2013}) would thus allow us to reach the  highest sensitivities also in the limit of low Rabi frequencies. The sensitivity could be further enhanced by improving the Rabi decay time $\tau_R$ using isotopically enriched diamond~\cite{Bala2009,Rondin2014} and by optimizing the photon collection efficiency using alternative tip geometries~\cite{Momenzadeh2015} or scanning probes made from [111] oriented diamond material~\cite{Neu2014}. In addition, the MW magnetic field sensitivity can be estimated from the full, quantitative field measurement (Fig. \ref{fig:fig3}) and is given by $\eta_{\text{meas}} = \delta B'_{\text{-,MW}} \sqrt{T}$, where $\delta B'_{\text{-,MW}}$ is the smallest measurable MW magnetic field, i.e. the fitting error to the Rabi fits, and $T$ the measurement time for each data point. For the values extracted from Fig. \ref{fig:fig3} we obtain a sensitivity of $\eta_{\text{meas}} = 15$ $\mu$T$/\sqrt{\text{Hz}}$~\cite{SI}, which is larger than the shot-noise limited sensitivity quoted above. This discrepancy is explained by the fact that for the measurement shown in Fig. \ref{fig:fig3}, we recorded full Rabi oscillations for each point of the scan, i.e. most of the data was taken for evolution times $\tau$, which do not yield optimal measurement sensitivities~\cite{SI}.\\


In conclusion, we have established scanning NV center spins as a valuable resource to sensitively detect and image MW magnetic fields on the nanoscale. Our results indicate an imaging resolution of $\sim25~$nm together with a shot noise limited MW magnetic field sensitivity of $680~$nT$/\sqrt{\text{Hz}}$, resulting in a sensitivity to the generating currents of few nA/$\sqrt{\text{Hz}}$, all at frequencies $\sim3~$GHz. Extending the bandwidth of detection to the range above $20$ GHz can be achieved by placing our microscope in a sufficiently strong magnetic field~\cite{Stepanov2015} and would have profound impact for applications in MW device characterization, as currently available field imaging techniques cannot operate in this frequency range~\cite{Sayil2005}. It should be noted that detection of microwave fields through Rabi oscillations has also been implemented for $^{87}$Rb vapor cells~\cite{Boehi2010,Boehi2012,Horsley2013,Affolderbach2015}. Such devices operate with tens of $\mu$m spatial resolution over a mm to cm detection window~\cite{Horsley2015}, compared to our nanoscale spatial resolution over a tens of $\mu$m detection window, and thus provide a complementary wide field imaging tool to our NV scanning magnetometers. Finally, recent experiments have demonstrated that spin wave excitations in nanomagnetic systems can be adressed via MW NV magnetometry~\cite{vanderSar2015}. There external DC magnetic fields are used to bring the spin wave excitation frequency into resonance with the NV spin transition and thus enables a detection of the spin wave amplitude via the NV Rabi frequency. 
Combining this detection scheme with our ability to image MW magnetic fields at nanoscale resolution would therefore form an exciting avenue that could allow for real space imaging of spin wave excitation in nanomagnets~\cite{Spinelli2014} or skyrmion core dynamics~\cite{Nagaosa2013}. 

\section*{Acknowledgments}
We gratefully acknowledge financial support through the NCCR QSIT, a competence center funded by the Swiss NSF, and through SNF Grant No. 200021 143697/1 and Grant No. 155845. This research has been partially funded by the European Commission's 7. Framework Program (FP7/2007-2013) under grant agreement number 611143 (DIADEMS). E.N. acknowledges funding via the NanoMatFutur program of the german ministry of education and and reasearch. We thank B. Shields, A. Barfuss, A. Horsley and P. Treutlein for fruitfull discussions.

%


\onecolumngrid

\section{Supporting Information}

\section{\label{Proj}Projection of the MW magnetic field on the NV axis and determination of $\sigma_{\pm}$ MW polarization}

As described in the manuscript, the MW current $\vec{I}_{MW}$ in the stripline generates a linearly polarized, MW magnetic field $\vec{B}_{\text{MW}}\left(\vec{r}\right) \cos{\left(\omega_{\text{MW}} t\right)}$ with a frequency $\omega_{\text{MW}}$ in the GHz range at the position of the NV spin $\vec{r}=\left(x,y,z\right)$. In the following we will assume a fixed frequency $\omega_{\text{MW}}$ and position in space $\vec{r}=\left(x,y,z\right)$. For simplicity, we neglect the frequency component and suppress the dependency on $\vec{r}$ in the notation. We have
\begin{equation}
\vec{B}_{\text{MW}} = \begin{pmatrix} B_{\text{x,MW}} \\ B_{\text{y,MW}} \\ B_{\text{z,MW}}\end{pmatrix}
\end{equation} 
The projection of the MW magnetic field on the NV axis is given by 
\begin{equation}
\begin{pmatrix} B'_{\text{x,MW}} \\ B'_{\text{y,MW}} \\ B'_{\text{z,MW}}\end{pmatrix} = \mathfrak{R}\left(\varphi,\theta\right) \cdot \begin{pmatrix} B_{\text{x,MW}} \\ B_{\text{y,MW}} \\ B_{\text{z,MW}}\end{pmatrix}
\label{Bproj}
\end{equation}
where $\mathfrak{R}\left(\varphi,\theta\right)$ corresponds to the transformation from the laboratory \{x,y,z\} to the reference frame \{x',y',z'\} defined by the NV axis  $\widehat{e}_{z}$. The transformation matrix holds
\begin{equation}
\mathfrak{R} \left(\varphi,\theta\right) = \begin{pmatrix} \cos{\varphi} \cos{\theta} & \sin{\varphi} \cos{\theta} & -\sin{\theta} \\ -\sin{\varphi} & \cos{\varphi} & 0 \\ \cos{\varphi} \sin{\theta} & \sin{\varphi} \sin{\theta} & \cos{\theta} \end{pmatrix}
\label{Rmatrix}
\end{equation}
where $\varphi$ is the angle between the NV spin axis $\widehat{e}_{z}$ and the MW current direction $\vec{I}_{\text{MW}}$ in the stripline plane (xy plane) and $\theta$ the angle with the surface normal of the stripline (z axis) (see Fig. 2a of the main text). Using Eq. \ref{Rmatrix} and \ref{Bproj} we have
\begin{equation}
\begin{pmatrix} B'_{\text{x,MW}} \\ B'_{\text{y,MW}} \\ B'_{\text{z,MW}}\end{pmatrix} = \begin{pmatrix} B_{\text{x,MW}} \cos{\varphi} \cos{\theta} + B_{\text{y,MW}} \sin{\varphi} \cos{\theta} - B_{\text{z,MW}} \sin{\theta} \\ -B_{\text{x,MW}} \sin{\varphi} + B_{\text{y,MW}} \cos{\varphi} \\ B_{\text{x,MW}} \cos{\varphi} \sin{\theta} + B_{\text{y,MW}} \sin{\varphi} \sin{\theta} + B_{\text{z,MW}} \cos{\theta} \end{pmatrix}
\label{Bprojfinal}
\end{equation}
The dependance on the orientation $\left(\varphi,\theta\right)$ of the NV spin has been omitted in the notation for the sake of simplicity.

The NV spin is coherently manipulated by the MW magnetic field component $\vec{B'}_{\bot,\text{MW}}$ orthogonal to the NV spin axis. This orthogonal component is linearly polarized and can be written as 
\begin{equation}
\vec{B'}_{\bot,\text{MW}} = \frac{1}{\sqrt{2}}\vec{B'}_{-,\text{MW}} + \frac{i}{\sqrt{2}}\vec{B'}_{+,\text{MW}}
\end{equation}
where $\vec{B'}_{-(+),\text{MW}}$ corresponds to the right (left) handed circular MW polarization component $\sigma_{-(+)}$. The amplitude of the right (left) handed polarization is defined as follows
\begin{equation}
B'_{\mp,\text{MW}} = |B'_{\text{x,MW}} \pm i B'_{\text{y,MW}}|
\label{sigma_pol}
\end{equation}
Reminder: The MW magnetic field polarization $\vec{B'}_{\mp,\text{MW}}$ depends on the position $\vec{r}=\left(x,y,z\right)$ and the orientation $\left(\varphi,\theta\right)$ of the NV spin, which has been omitted in the notation for the sake of simplicity.

\section{Spatial distribution of MW magnetic field: finite element simulation}

In the following section, we compute the MW magnetic field using a finite element simulation (RF module, COMSOL Multiphysics).

\begin{figure}[hbt!]
\includegraphics[width = 8 cm]{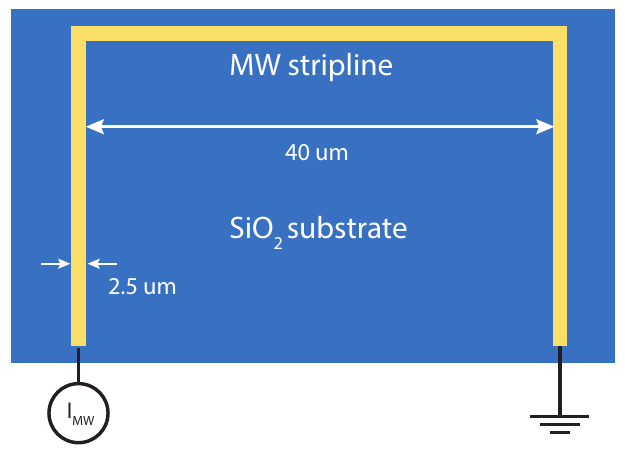}
\caption{Stripline geometry used for finite element simulation}
\label{figS4}
\end{figure}

We assume a 50 $\Omega$ Pd stripline (width $w = 2.5 $ $\mu$m, thickness $t = 60$ nm) patterned on an undoped Si substrate covered by 300 nm of SiO$_2$ surrounded by air. A current signal $I_{\text{MW}} = I_p \sin{\omega_{\text{MW}} t}$ is applied at the input of the stripline, while the other end is being grounded (Fig. \ref{figS4}a). We use a frequency $\omega_{\text{MW}} = 2.825$ GHz and an excitation current of $I_p = 10$ mA, which corresponds to a MW power of $P_{\text{RF}} = 7$ dBm. The latter is close to the estimated power used in the experiment.

We first compute the MW current density $\vec{J}_{\text{MW}}\left(\vec{r}\right)$ at each position $\vec{r}$ in the laboratory frame for the given frequency $\omega_{\text{MW}}$ (Fig. \ref{figS4}a). The MW current density is homogeneous in the MW stripline, except at the right angled bends of the conductor. It should be noted that the skin effect can be neglected, as the skin depth of palladium ($\delta = 3.5$ $\mu$m at 2.825 GHz) is larger than the width and thickness of the MW stripline. 

\begin{figure}[hbt!]
\includegraphics[width = 16 cm]{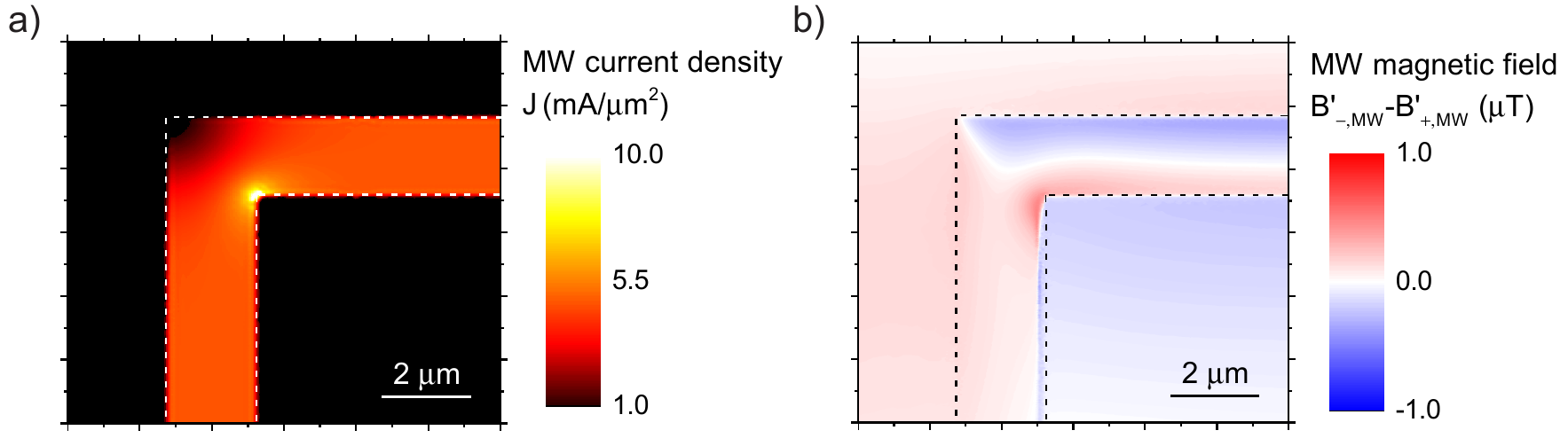}
\caption{a) MW current density and b) difference $B'_{-,MW}-B'_{+,MW}$ obtained from finite element simulations with COMSOL Multiphysics}
\label{figMWsim}
\end{figure}

The MW magnetic field $\vec{B}_{\text{MW}}\left(\vec{r}\right)$ can then be calculated from the current density and transformed into a $\sigma_-$ MW polarization using Eq. \ref{Bprojfinal} and \ref{sigma_pol}. The obtained $\sigma_-$ polarization of the MW magnetic field is in excellent agreement with the 1D profiles in Fig. 3b and c of the main text. Also the simulated 2D MW field distribution in a xy plane 25 nm above the stripline (Fig. 3d main text) is in good agreement with the experiment (Fig. 3a main text). Differences might arise from a non ideal stripline and eventual impedance mismatch between the MW stripline and the connecting RF cables in the experiment.

Finally, it should be noted that the simulations revealed almost identical field distributions for $\sigma_-$ and $\sigma_+$ polarization. The difference $B'_{-,MW}-B'_{+,MW}$, depicted in Figure \ref{figMWsim}b in an xy plane 25 nm above the surface of the stripline, is on the order of a few hundred nT and therefore smaller than the smallest detectable MW field (see last section of the supporting material). The MW magnetic field is therefore mostly linearly polarized for this particular stripline geometry and consequently we did not measure the $\sigma_+$ polarization in our experiment. However, we would like to emphasize that such a measurement is straightforward and can in principle be performed by tuning the sensing frequency to $|0\rangle \rightarrow |+ 1\rangle$ transition of the NV center as described in the main text.

\section{Spatial distribution of MW magnetic field: analytical model}
\subsection{Magnetic vector potential and field of a stripline}
In the following we provide an analytical expression for the MW magnetic field generated by the MW current in a stripline, which has a thickness $t = 60$ nm much smaller than its width $w = 2.5$ $\mu$m. The stripline can be considered as an infinitely thin wire and according to the finite element simulation in the previous section we can assume a homogeneous MW current density, except at the right angled bend of the MW conductor. For the straight part of the MW conductor aligned for example along the y direction, we therefore have a homogeneous current density $\vec{J} = \left(0,J_y=J,0\right)$ and the problem can be treated in the two dimensional xz plane. We neglect also the substrate and therefore assume an infinitely thin MW stripline in vacuum. The NV spin is located at a position $\left(x,d+t/2\right)$. 

\begin{figure}[hbt!]
\includegraphics[width = 8 cm]{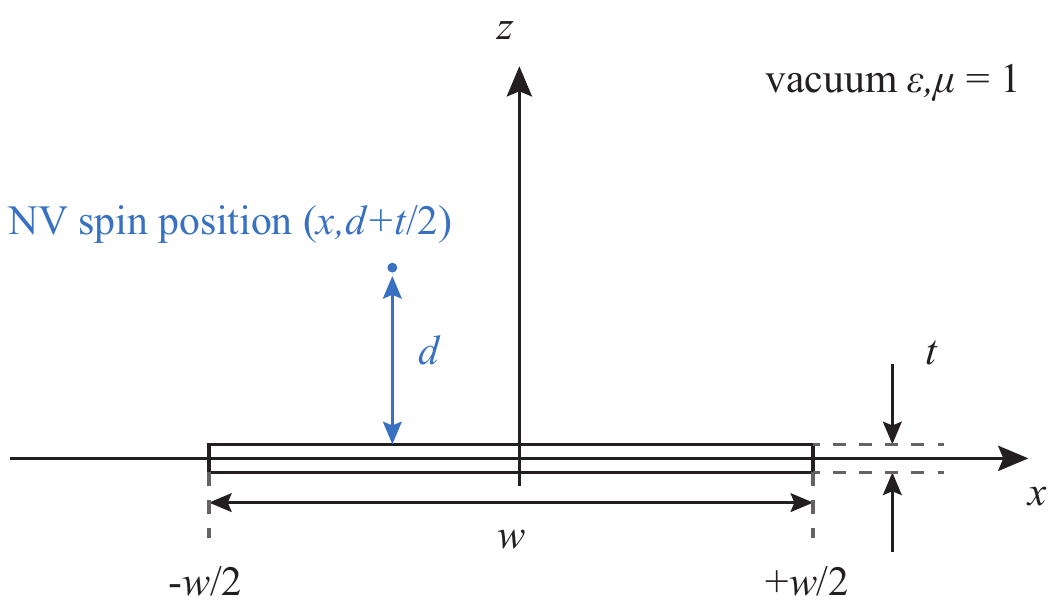}
\caption{}
\label{figS1}
\end{figure}
From Poisson's equation, we calculate the magnetic vector potential oriented in y direction
\begin{equation}
A_{y}(\vec{r},J) = \frac{\mu_0}{4 \pi} \int{\frac{J\left(x',z'\right)dV'}{|\vec{r}-\vec{r}'|}}
\label{magn_pot}
\end{equation}
where $\vec{r}=(x,d+t/2)$ are the coordinates of the NV spin where $A_{y}$ is evaluated and $\vec{r}'=(x',z')$ the coordinates of the MW current density in the wire. 

Equation \ref{magn_pot} can be evaluated in analogy to an electrostatic potential with a charge distribution $\rho\left(x',z'\right)$
\begin{equation}
\Phi = \frac{1}{4 \pi \epsilon_0} \int{\frac{\rho \left(x',z'\right)dV'}{|\vec{r}-\vec{r}'|}} = -\frac{1}{2 \pi \epsilon_0} \iint{\rho\left(x',z'\right)ln\left(\frac{\sqrt{(x-x')^2+(z-z')^2}}{r_0}\right)dx'dz'}
\label{elec_pot}
\end{equation}
where $r_0$ is an integration constant. Consequently Eq. \ref{magn_pot} therefore becomes
\begin{equation}
A_{y}(\vec{r},J) = -\frac{\mu_0}{2 \pi} \iint{J\left(x',z'\right)ln\left(\frac{\sqrt{(x-x')^2+(z-z')^2}}{r_0}\right)dx'dz'}
\end{equation}
Following the assumption of the infinitely thin wire, the integration over $z'$ amounts to $J \cdot t$. With integration boundaries as depicted in Fig. \ref{figS1} the magnetic potential reduces to
\begin{equation}
A_{y}(\vec{r},J) = -\frac{\mu_0 J t}{2 \pi} \int_{-w/2}^{w/2}{ln\left(\sqrt{(x-x')^2+z^2}\right)dx'}
\label{Afinal}
\end{equation}
The corresponding MW magnetic field is determined from
\begin{equation}
\vec{B}_{\text{MW}}(\vec{r},J) = - \widehat{e}_y \times \nabla A_y (\vec{r},J) = \begin{pmatrix} -\partial_z A_y (\vec{r},J) \\ 0 \\ \partial_x A_y (\vec{r},J)\end{pmatrix}
\label{Bfinal}
\end{equation}

\subsection{MW magnetic field projection on NV axis $\vec{B}'_{\text{MW}}(\vec{r},J,\varphi,\theta)$ and determination of $\sigma_-$ polarization component $B'_{-,\text{MW}}(\vec{r},J,\varphi,\theta)$}
The MW magnetic field projected on the NV axis can be calculated by inserting Eq. \ref{Bfinal} into Eq. \ref{Bprojfinal} and yields
\begin{equation}
\vec{B}'_{\text{MW}}(\vec{r},J,\varphi,\theta) = \begin{pmatrix} -\partial_z A_y (\vec{r},J) \cos{\varphi} \cos{\theta} - \partial_x A_y (\vec{r},J) \sin{\theta} \\ \partial_z A_y (\vec{r},J) \sin{\varphi}\\ -\partial_z A_y (\vec{r},J) \cos{\varphi} \sin{\theta} + \partial_x A_y (\vec{r},J) \cos{\theta} \end{pmatrix}
\label{Bprojfinal_anal}
\end{equation}
With Eq.\ref{sigma_pol} and \ref{Bprojfinal_anal}, one can finally determine the polarization component $\sigma_-$
\begin{eqnarray}
B'_{-,\text{MW}}(\vec{r},J,\varphi,\theta) &=& |B'_{\text{x,MW}}(\vec{r},J,\varphi,\theta) + i B'_{\text{y,MW}}(\vec{r},J,\varphi,\theta)| \nonumber \\ &=&  |-\partial_z A_y (\vec{r},J) \cos{\varphi} \cos{\theta} - \partial_x A_y (\vec{r},J) \sin{\theta} + i \partial_z A_y (\vec{r},J) \sin{\varphi}|
\label{fit_anal}
\end{eqnarray}
Considering $\vec{r}=(x,d+t/2)$, the MW magnetic field polarization depends on $x,d,t,J,\varphi,\theta$. With the thickness being $t = 60$ nm and $x$ the dependant variable, Eq. \ref{fit_anal} can be used as a fitting function were the distance $d$, the MW current density $J$, and the angles $\varphi$ and $\theta$ are free (fit) parameters. As demonstrated in the main text, fitting the experimental data with Eq. \ref{fit_anal} shows an excellent agreement, confirming our initial assumptions that the substrate and the thickness of the stripline can be neglected. Finally, it should be noted that this analytical model is only valid under the assumption of a homogeneous current distribution and would not be accurate, for instance across the sharp bend of our MW conductor.

\subsection{Determination and uncertainties of the fit parameters $d,J,\varphi,\theta$}

The MW magnetic field amplitude $B'_{-,\text{MW}}$ at a position $\vec{r}$ is determined in Fig. 3 of the main text by fitting the Rabi oscillations measured at this position (see main text). The corresponding Rabi fit function is given by
\begin{equation}
F(\gamma_{\text{NV}} B'_{-,\text{MW}} \tau) = F_0(1-C/2) + C F_0/2 \cdot \cos\left({2 \pi \gamma_{\text{NV}} B'_{-,\text{MW}} \tau} \right) \cdot e^{-\tau / \tau_\text{R}}
\end{equation}
where $\tau_{\text{R}}$ is the Rabi decay time, $\gamma_{\text{NV}}$ the gyromagnetic ratio of the NV, $\tau$ the MW pulse duration and $C$ a factor related to the contrast and $F_0$ the bare fluorescence. The MW magnetic field $B'_{-,\text{MW}}$ yields a fit error $\delta B'_{\text{-,R}}$, depicted by the error bars in Fig. 3 of the main text. In a second step, the resulting MW magnetic field profile is fitted with the analytical function $B'_{-,\text{MW}}(d,J,\varphi,\theta)$ of Eq. \ref{fit_anal}. This fit gives the values $p_i$ for each parameter $\{d,J,\varphi,\theta\}$ with the corresponding fit error $\delta p_{\text{i,fit}}$.

The uncertainty of each $p_i$ is given by (i) the uncertainty of the Rabi fit $\delta p_{\text{i,R}}$ and (ii) the uncertainty $\delta p_{\text{i,fit}}$ of the fit with $B'_{-,\text{MW}}(d,J,\varphi,\theta)$ and yields
\begin{equation}
\delta p_i = \sqrt{\delta p_{\text{i,R}}^2+\delta p_{\text{i,fit}}^2}
\end{equation} 
In order to account for the uncertainty arising from the Rabi fit $\delta p_{\text{i,R}}$, we fit the dataset $B'_{-,\text{MW}} + \delta B'_{\text{-,R}}$ and $B'_{-,\text{MW}} - \delta B'_{\text{-,R}}$ with Eq. \ref{fit_anal}. The fit then yields a parameter set $p_{i,+}$ and $p_{i,-}$, respectively. The uncertainty is given by
\begin{equation}
\delta p_{i,R} = \frac{p_{i,+}-p_{i,-}}{2}
\end{equation} 
The uncertainty $\delta p_{\text{i,fit}}$ can be directly extracted from the fit with Eq. \ref{fit_anal}. Tables \ref{fig4a} and \ref{fig4b} show the values and uncertainties of all parameters extracted in Fig. 3b and c of the main text. 
\begin{table}[hbt!]
	\centering
		\begin{tabular}{|c||c||c|c||c|}\hline
		parameter &$p_i$ & $\delta p_{\text{i,R}}$ & $\delta p_{\text{i,fit}}$ & $\delta p_{\text{i}}$ \\\hline\hline
		d & 25 nm & 2 nm & 5 nm & 5 nm\\\hline
		J & 3.9 mA$/\mu$m$^2$ & 0.05 mA$/\mu$m$^2$ & 0.03 mA$/\mu$m$^2$ & 0.05 mA$/\mu$m$^2$ \\\hline
		$\theta$ & $49^{\circ}$ & $0.6^{\circ}$ & $0.5^{\circ}$ & $1^{\circ}$		 \\\hline
		$\phi$ & $104^{\circ}$ & $2.2^{\circ}$ & $1^{\circ}$ & $2^{\circ}$		 \\\hline
		\end{tabular}
		\caption{Summary of fit parameters including uncertainties for Fig. 3b of the main text}
		\label{fig4a}
\end{table}
\begin{table}[hbt!]
	\centering
		\begin{tabular}{|c||c||c|c||c|}\hline
		parameter &$p_i$ & $\delta p_{\text{i,R}}$ & $\delta p_{\text{i,fit}}$ & $\delta p_{\text{i}}$ \\\hline\hline
		d & 64 nm & 6 nm & 5 nm & 8 nm\\\hline
		J & 4.05 mA$/\mu$m$^2$ & 0.06 mA$/\mu$m$^2$ & 0.02 mA$/\mu$m$^2$ & 0.05 mA$/\mu$m$^2$ \\\hline
		$\theta$ & $48^{\circ}$ & $1.3^{\circ}$ & $0.5^{\circ}$ & $1^{\circ}$		 \\\hline
		$\phi$ & $11^{\circ}$ & $0.3^{\circ}$ & $0.8^{\circ}$ & $1^{\circ}$		 \\\hline
		\end{tabular}
		\caption{Summary of fit parameters including uncertainties for Fig. 3c of the main text}
		\label{fig4b}
\end{table}

\begin{figure}[hbt!]
\includegraphics[width = 16 cm]{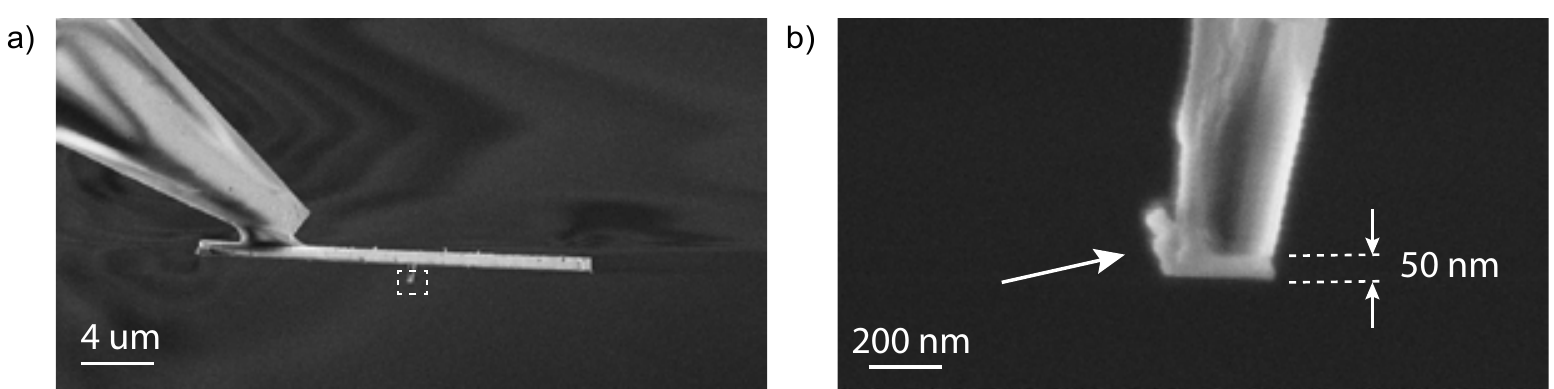}
\caption{NV scanning probe after an experiment}
\label{figScanti}
\end{figure}

As discussed in the main text, the difference of $\varphi$ between Tables \ref{fig4a} and \ref{fig4b} is fully consistent with the $90^{\circ}$ rotation of the MW current with respect to the scanning probe and the NV spin. The difference in the distance $d$ can however be attributed to a spacing layer between the stripline and the scanning probe. Indeed, we observed the presence of a spacing layer at the end of the pillar on several diamond scanning probe used in magnetometry experiment (Fig. \ref{figScanti}). The layer has a typical thickness of a few tens of nanometers and is most likely composed of organic material, for example residual resist or dust particles picked up somewhere on the sample surface during an AFM scan. It should be noted that metallic particles would lead to quenching of the NV fluorescence~\cite{Novotny2006}. As we found no evidence for fluorescence quenching, we can therefore exclude a metallic spacing layer for the scanning probe used in this experiment.

\subsection{MW magnetic field $\sigma_-$ polarization $B'_{-,\text{MW}}(\vec{r},J,\varphi,\theta)$ dependance on distance and orientation of NV spin}

In this section, we illustrate the dependance on the distance and the NV spin orientation. We calculate the MW magnetic field $B'_{-,\text{MW}}(\vec{r},J,\varphi,\theta)$ for different sets of parameters. As the MW current density is directly proportional to the MW magnetic field, we assume $J=4$ mA/$\mu$m$^2$ as determined in Fig. 3 of the main text. 
\begin{figure}[hbt!]
\includegraphics[width = 8 cm]{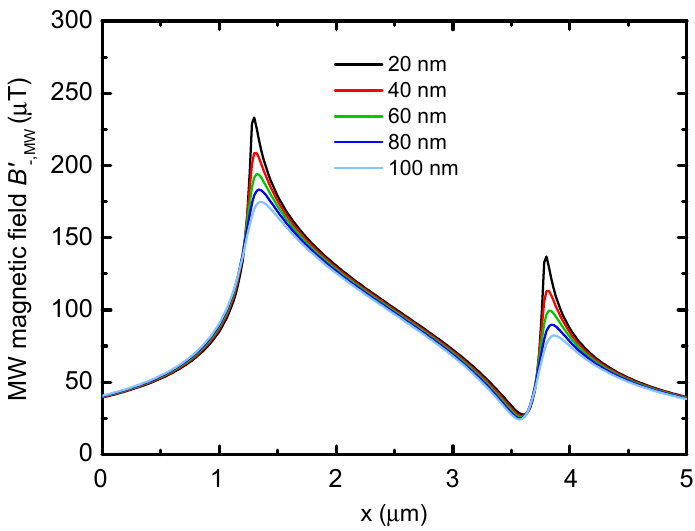}
\caption{NV spin distance dependance of the MW magnetic field: the profiles are calculated from Eq.\ref{fit_anal} for different distances $d$ using the fit parameters from Fig. 3b of the main text $J=4$ mA/$\mu$m$^2$, $\theta=49^{\circ}$ and $\varphi=104^{\circ}$.}
\label{figS2}
\end{figure}
Fig. \ref{figS2} shows the evolution of the MW magnetic field profile along the x direction for different distance $d$ of the NV spin to the stripline. As the distance $d$ decreases, sharp peaks develop in the MW magnetic field close to the edge of the stripline, which can be understood in terms of stronger MW magnetic field gradients on the edge.

\begin{figure}[hbt!]
\includegraphics[width = 16 cm]{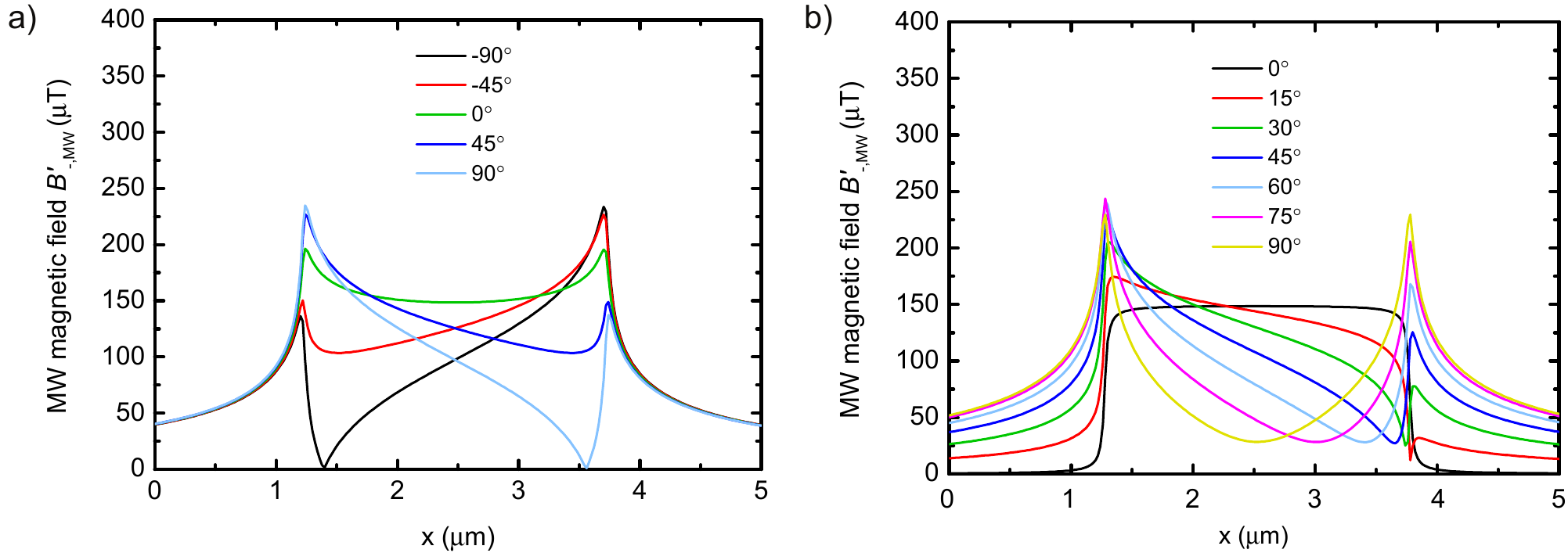}
\caption{NV spin orientation dependance of the MW magnetic field: (a) MW magnetic field profiles calculated from Eq. \ref{fit_anal} for different in plane angles $\varphi$ (with respect to the MW current) using the fit parameters from Fig. 3b of the main text $J=4$ mA/$\mu$m$^2$, $\theta=49^{\circ}$ and $d = 25$ nm. (b)  MW magnetic field profiles calculated from Eq. \ref{fit_anal} for different out of plane angles $\theta$ (with respect to the MW current) using the fit parameters from Fig. 3b of the main text $J=4$ mA/$\mu$m$^2$, $\varphi=104^{\circ}$ and $d = 25$ nm.}
\label{figS3}
\end{figure}

In contrast to the distance $d$ which only causes small modifications of the MW magnetic field close to the edge of the stripline, the orientation of the NV spin has a dramatic impact on the MW magnetic field profile (Fig. \ref{figS3} and Fig. 3 of the main text).

\section{MW magnetic field sensitivity}
\subsection{Photon shot noise limited sensitivity}
Moreover, we estimate the photon shot noise limited MW magnetic field sensitivity $\eta_{\text{photon}}$. This sensitivity is given by 
\begin{equation}
\eta_{\text{photon}} = \delta B'_{\text{photon}} \sqrt{T}
\end{equation}
where $\delta B'_{\text{photon}}$ is the smallest MW magnetic field obtained for a signal-to-noise ratio of $1$ and $T$ the measurement time. In our experiment, the signal is defined as 
\begin{equation}
S = F_0(1-C/2) + C F_0/2 \cdot \cos\left({2 \pi \gamma_{\text{NV}} \delta B'_{\text{photon}} \tau} \right) \cdot e^{-\tau / \tau_\text{R}}
\label{Rabi}
\end{equation}
where $\tau_{\text{R}}$ is the Rabi decay time, $\gamma_{\text{NV}}$ the gyromagnetic ratio of the NV, $\tau$ the MW pulse duration and $C = (F_0-F_1)/F_0$ the fluorescence contrast between the $|0\rangle$ and the $|1\rangle$ state and $F_0$ ($F_1$) the fluorescence (in counts per shot) in the $|0\rangle$ state ($|1\rangle$ state). For small values of the accumulated phase $\Phi = 2 \pi \gamma_{\text{NV}} \delta B'_{\text{photon}} \tau$~\cite{Taylor2008}, the signal can be approximated by
\begin{equation}
S \approx 2 \pi \gamma_{\text{NV}} \delta B'_{\text{photon}} \tau\frac{C F_0}{2} e^{-\tau / \tau_\text{R}}
\end{equation}
In order to obtain maximum signal and thus optimal sensitivity, the signal has to be evaluated at $\tau = \tau_{\text{R}}/2$~\cite{Rondin2014}. For a signal to noise ratio of 1 and assuming photon shot noise of $\sqrt{\text{F}_0}$ a minimal field of
\begin{equation}
\delta B'_{\text{photon}} = \frac{2 \sqrt{e}}{\pi \gamma_{\text{NV}} C \sqrt{\text{F}_0} \tau_{\text{R}} } 
\end{equation}
can be calculated.

It should be noted that the measurement time for a single shot experiment is equivalent to $\tau = \tau_{\text{R}}/2$. During a total measurement time of $T$, $T/ \tau = 2 T/\tau_{\text{R}}$ repetitions of a single shot measurement can thus be performed. In this case, the fluorescence $F_0$ (in counts per shot) scales to $2 F_0 T/\tau_{\text{R}}$ and one finally obtains the shot noise limited sensitivity
\begin{equation}
\eta_{\text{photon}} = \delta B'_{\text{photon}} \sqrt{T} = \frac{2e \sqrt{\text{T}}}{\pi \gamma_{\text{NV}} C \sqrt{2 \text{F}_0T/\tau_{\text{R}}} \tau_{\text{R}}} = \frac{\sqrt{2e}}{\pi \gamma_{\text{NV}} C \sqrt{\text{F}_0 \tau_{\text{R}}} } 
\end{equation}
The value for $C$, $F_0$ and $\tau_{\text{R}}$ are extracted from the Rabi fit (equation. \ref{Rabi}) to the data in Fig. 1c of the main text and yield $C = 0.095$, $F_0 = 0.045$ counts/shot and $\tau_{\text{R}} = 4$ $\mu$s. The sensitivity therefore amounts to $\eta_{\text{photon}} = 680$ nT$/\sqrt{\text{Hz}}$. 

\subsection{Measured sensitivity}
The MW magnetic field sensitivity in our experiments is estimated from the quantitative MW magnetic field measurement in Fig. 3 of the main text. The sensitivity is defined by
\begin{equation}
\eta_{\text{meas}} = \delta B'_{\text{-,MW}} \sqrt{T}
\end{equation}
where $\delta B'_{\text{-,MW}}$ is the smallest measurable MW magnetic field, i.e. the fitting error to the Rabi frequency and thus the MW magnetic field and $T$ the measurement time for each data point. We obtain an average fitting error  $\delta B'_{\text{-,MW}} = 2$ $\mu$T of for both the image in Fig. 3a and the linecuts in Fig. 3b and c. With a measurement time of $T=67$ s we get a sensitivity $\eta_{\text{meas}} = 15$ $\mu$T$/\sqrt{\text{Hz}}$, which is $\sim 20$ times larger then the photon shot noise limited sensivitity. 

It should be noted that the shot noise limited sensitivity is calculated at the point of optimal signal to noise ratio during a Rabi oscillation measurement. Therefore, by measuring and fitting a complete Rabi oscillation we only obtain an averaged and lower overall sensitivity.

\end{document}